\documentclass[twocolumn,showpacs,showkeys]{revtex4}

\usepackage{epsfig}
\usepackage{amsmath}
\usepackage{bm}% bold math

\begin{document} 

\title{Phase diagram as a function of temperature and magnetic field   %
for magnetic semiconductors.}

\author{I. Gonz\'alez}
\email{faivan@usc.es}
\author{J. Castro}
\author{D. Baldomir}

\affiliation{Departamento de F\'{\i}sica Aplicada, Universidade        %
de Santiago de Compostela,\\ E-15706 Santiago de Compostela, Spain.}

\begin{abstract}
Using an extension of the Nagaev model of phase separation (E. L.      %
Nagaev, and A. I. Podel'shchikov, Sov. Phys. JETP, 71 (1990) 
1108), we calculate the phase diagram for degenerate
antiferromagnetic semiconductors  in the $T-H$ plane for different
current carrier densities. Both, wide-band semiconductors  and  
``double-exchange'' materials, are investigated.
\end{abstract}

%Add the option ``showpacs'' into documentclass to show them.

%Add the option ``showkeys'' into documentclass to show them.

\maketitle

\section{Introduction}

Degenerate antiferromagnetic semiconductors are obtained by            %
strongly doping antiferromagnetic semiconductors (e.g., europium 
chalcogenides or lanthanium manganites). Over a concentration range 
of doping impurities, the ground state of these compounds will be a 
mixture of antiferromagnetic (AF) and ferromagnetic (FM) phases
\cite{Nag83,Nag72,Kas73}. 
Starting from a pure compound and doping, we find that the ground state
at $T=H=0$ is AF and non-conducting up to some conduction electron
concentration driven by impurities, $n_{\text{d}}$, from which the
homogeneous state turns out to be unstable against phase separation.
The ground state becomes then inhomogeneous with a simply connected
AF phase and a multiply connected FM phase. The exact geometry of the 
multiply connected phase is of fractal nature. On increasing doping, a 
geometric transition takes place in which the topology of the sample 
changes due to percolation of the FM phase. Now, the ground state 
corresponds to a multiply connected AF phase and a simply connected FM 
one. The doping concentration at which this transition occurs is 
denoted by $n_{\text{T}}$. At this point an important change in the
conductivity is expected, the material becomes a conductor. On a
further increase of doping, a certain concentration, $n_{\text{u}}$, 
exits at which this phase-separated state starts to be unstable and 
the sample becomes again homogeneous but now FM.
The physical reason for having a phase-separated state in an           %
antiferromagnetic semiconductor at $T=H=0$ is the dependence of the
energy of the charge carriers that appear on doping on the magnetic 
order of the lattice. The charge carrier energy is lower if they move 
in a FM region. Therefore, by interaction with the spin system in the 
lattice, they are able to change its magnetic order, creating FM 
micro-regions and become trapped in them. In a degenerate magnetic
semiconductor this is a cooperative phenomenon: a number of electrons
are self-trapped in the same FM micro-region diminishing the energy 
per electron necessary to create it. When the carrier concentration 
is high enough, the phase-separated state turns out to be stable.   
For $T,\,H\neq0$ the free energy has different expressions for the 
AF and the FM parts. Because of this, a change in the temperature 
or the magnetic field produces a variation of the relative volume 
of the AF and FM parts. Therefore, a temperature- or magnetic 
field-induced percolation can occur, resulting in a complicated phase
diagram. 

In this article, we calculate the phase diagram in the $T-H$ plane for %
magnetic semiconductors within the doping range
$n_{\text{d}}<n<n_{\text{T}}$. The phase diagram is expected to be 
divided into three different stability regions, those corresponding 
to: insulating phase-separated state, conductive phase-separated state
and homogeneous state. Our calculations are based in a variational
principle for the free energy of the system developed by Nagaev and
Podel'shchikov in the reference \cite{Nag90} that generalizes the 
approach of the references \cite{Nag83,Nag72,Kas73} to the case of 
finite temperature.

\section{Calculation of the free energy}

In this section, we briefly review the variational method used.        %
More detailed calculations can be found in references
\cite{Nag83,Nag72,Kas73,Nag90}.
The microscopic description of the sample is provided by the 
Hamiltonian of the generalized Vonsovsky s-d model. The main 
parameters in this model are $W=2zt$, the carriers band-width,
$AS$ the exchange energy between conduction electrons and magnetic 
ions (s-d exchange energy), and $zIS^2$ the exchange energy between 
magnetic ions (d-d exchange energy). $I<0$ is the exchange integral
between first-nearest neighbors magnetic ions. The smallest parameter
is the d-d exchange energy. We differentiate two possibilities
depending on the relative value of $W$ and $AS$. In the case $W>>AS$,
we have a wide-band semiconductor (e.g. $EuTe$). In the opposite case 
$W<<AS$, we talk about a ``double-exchange'' material (e.g. lanthanium 
manganites). For the sake of definiteness the sign of $A$ is assumed
to be positive. 

The free energy is expresed as a function of two variational
paramaters: the ratio $x$ of the volumes of the AF and FM phases,
$x=\frac{V_{A}}{V_{F}}$, and $R$, the radius of the spheres which form 
the multiply-connected phase. All of the electrons are assumed to be 
found in the FM part of the phase-separated state \cite{Art1}.
Although this is not strictly true at $T\neq0$, it is a good 
approximation because the number of electrons in the AF part is 
proportional to $e^{{\frac{-U}{T}}}$, where $U\sim W$ if $W>>AS$ or
$U\sim AS$ if $W<<AS$, and in both cases we have $U>>T$.
The trial free energy is given by the expression:

\begin{equation}
F=E_{\text{kin}}+E_{\text{sur}}+E_{\text{C}}+F_{\text{mag}}
\label{energy}
\end{equation}
where $E_{\text{kin}}$ is the standard bulk kinetic energy of the      %
conduction electrons, $E_{\text{sur}}$ is the surface electron energy, 
i.e. the correction to  $E_{\text{kin}}$ that appears due to the
quantization of the electron motion in regions of finite dimensions
\cite{Bal70}. In view of the degeneracy of the electron gas, the
contribution of the thermal excitations to its free energy can be 
neglected. $E_{\text{C}}$ is the electrostatic energy  due to the
inhomogenous electronic density and $F_{\text{mag}}$ is the magnetic 
free energy calculated in the mean field approximation. Due to the 
absence of conduction electrons on the AF part of the phase-separated
state, this trial free energy is valid for both $W>>AS$ and $W<<AS$. 

The expressions for these quantities are:

\begin{eqnarray}
E_{\text{kin}}&=&\frac{3}{5}
\mu\left(n\right)n\left(1+x\right)^{\frac{2}{3}}\nonumber\\
E_{\text{sur}}&=&
\frac{3}{16}\left(\frac{\pi}{6}\right)^{\frac{1}{3}}\frac{\beta}{R}
\mu\left(n\right)n^{\frac{2}{3}}\left(1+x\right)^{\frac{1}{3}}\nonumber\\
E_{\text{C}}&=&\frac{2\pi}{5\epsilon_{r}}\frac{e^{2}n^2 R^{2}}{a}f(x)
\end{eqnarray}

\begin{widetext}
\begin{eqnarray}
f(x)=
\begin{cases}
2x+3-3\left(1+x\right)^{\frac{2}{3}} 
& \text{if FM spheres in an AF matrix}, \\
x\left(2+3x-3x^{\frac{1}{3}}\left(1+x\right)x^{\frac{2}{3}}\right)
& \text{if AF spheres in a FM matrix}.
\end{cases}
\end{eqnarray}
\end{widetext}
where $\mu\left(n\right)=t\left(6\pi^{2}n\right)^{\frac{2}{3}}$ and    %
$\beta=3$ if the FM part is multiply connected, or $\beta=3x$ if  it is
simply connected. The Coulomb energy per unit cell is found by 
separation of the crystal into  Wigner cells. This method provides a 
good approximation to the electrostatic energy for small volumes of the 
alien phase, but it supposes to admit that the crystal is isotropic and
homogeneous regarding the spatial distribution of these spheres. So we
cannot describe the fractal nature of the boundary that separes both 
phases. 

The free energy of the magnetic system is calculated using the mean    %
field approximation. It is assumed that in the FM part conduction 
electrons are completely polarized. Then: 

\begin{equation}\label{ meanfield}
F_{\text{mag}}=\frac{x}{1+x}F_{\text{mag}}(H,T) +
\frac{1}{1+x}F_{\text{mag}}(H+H_{\text{e}},T)
\end{equation}
where:

\begin{widetext}
\begin{eqnarray}
F_{\text{mag}}(H,T)=
\begin{cases}
-\frac{H^{2}}{4J}+\frac{1}{2}JS_{1}^{2}-
T\ln\left[\sum_{m=-S}^{S}\exp\left(\frac{mJS_{1}}{T}\right)\right] 
& \text{if $H\le2JS_{1}$},\\[1.5ex]
-\frac{1}{2}JS_{2}^{2}-
T\ln\left[\sum_{m=-S}^{S}\exp\left(\frac{H-mJS_{2}}{T}\right)\right]
& \text{if $H\ge2JS_{1}$}.
\end{cases}
\end{eqnarray}
\end{widetext}
and the values $S_{1}, S_{2}$ verify the following equations:

\begin{eqnarray}
S_{1}&=&SB_{S}\left(\frac{SJS_{1}}{T}\right)\nonumber\\[1.5ex]
S_{2}&=&SB_{S}\left(\frac{H-SJS_{2}}{T}\right)
\end{eqnarray}
where $B_{S}$ is the Brillouin function. Moreover, we have defined 
$H_{\text{e}}=\frac{ASn\left(1+x\right)}{2}$ as the mean field created
by the conduction electrons on the localized d-spin system. The 
preceding energies where obtained taking into account that in the 
case of small fields, $H\le2JS_{1}$, the localized d-spins system is 
in a non-collinear AF (NCAF) state, being $S_{1}$ the mean d-spin value
in each of the sub-lattices. In the case of high fields, $H\ge2JS_{1}$, 
the sample goes to a non-saturated FM (NSFM) state, being $S_{2}$ the
mean value of the localized d-spin per site \cite{Nag83}. 

\section{Results.}

The stationary state of the system is determined from the condition    %
that the total free energy of the system, equation (\ref{energy}), be
a minimum with respect to the  variational parameters $x$ and $R$. The 
parameter $R$ only appears in the surface and Coulomb energies. 
Minimizing the sum $Q=E_{\text{sur}}+E_{\text{C}}$ with respect to
$R$ for fixed $x$ leads to the following expression:

\begin{equation}
Q_{\text{min}}=\alpha\gamma\beta(\frac{f}{\beta})^{\frac{1}{3}}
\label{Qmin}
\end{equation}
where the following definitions are used:
\begin{eqnarray}
\alpha&=&3\left(\frac{2\pi}{5}\right)^{\frac{1}{3}}
\left(\frac{9}{32}\right)^{\frac{2}{3}}
\left(\frac{\pi}{6}\right)^{\frac{2}{9}}\approx1.2 \nonumber\\
\gamma&=&\left(\frac{n^{\frac{1}{3}}e^{2}}
{\epsilon_{r}a}\mu^{2}n\right)^{\frac{1}{3}}
\end{eqnarray}

After substitution of equation (\ref{Qmin}) in equation (\ref{energy}) %
minimization with respect to $x$ is carried out numerically.

For $n<n_{\text{T}}$ three different phases are expected: insulating 
phase-se\-pa\-rated, conducting phase-separated and homogeneous states. 
We calculate the regions of absolute stability of each phase in the 
plane $T-H$ for different conduction electron concentrations in the 
sample. The boundary lines between phases are calculated numerically. 
Below we present these diagrams for different concentrations, all 
below $n_{\text{T}}$, for both wide-band \cite{Nag90} and
``double-exchange'' materials. 

Our results indicate that for both kinds of materials, all of the      %
phase transitions between homogeneous and phase-separated states are 
first-order. Preliminary calculations including the possibility of
spin waves confirm this result \cite{Nag00B}. The transition between
insulator and conductive phase-separated states is found close to 
second-order. The jump in the parameter $x$ at the transition is about
$20\%$ at $T=0$, growing with the transition temperature. This small
jump in the parameter does not permit elucidating whether in a real 
sample the transition is of first- or second-order. We have to stress
that the boundary between the separate states is of a fractal nature, 
and our approximation, that considers it as homogeneous and 
isotropically distributed spheres, does not work near the transition 
point. In any case, due to the smallness of the jump in $x$ our 
approximation provides accurate values of $T$ and $H$ at the
transition point. 

\subsection{$W>>AS$}

\begin{figure}
%\begin{figure*}
\epsfig{file=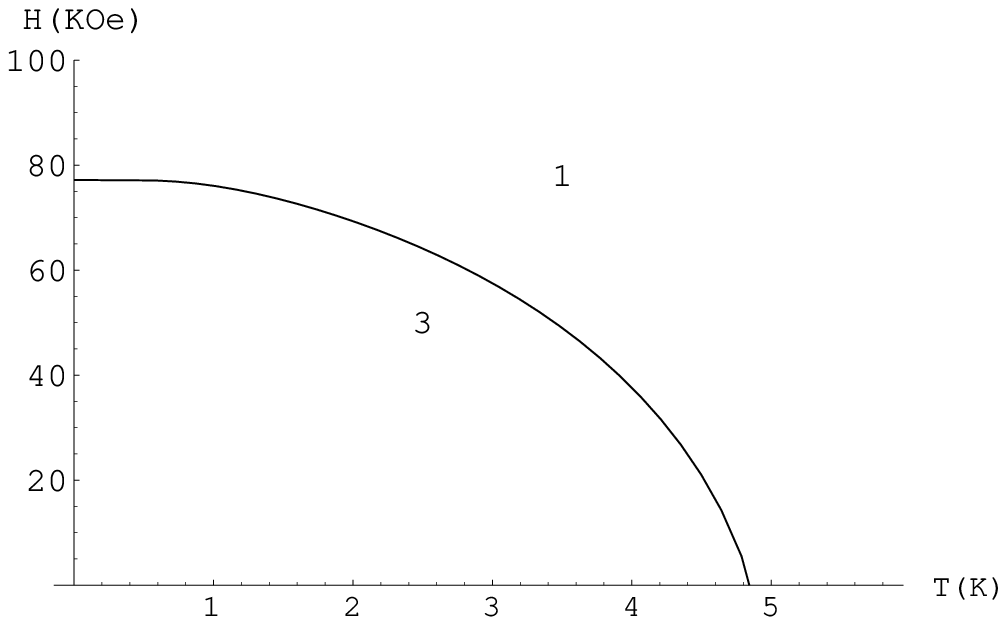,height=4.2 cm}
\caption{\label{fig1}Phase diagram for a wide-band AFS
with carrier concentration $n=0.10\, n_{\text{T}}$} 
%\end{figure}

%\begin{figure}
\epsfig{file=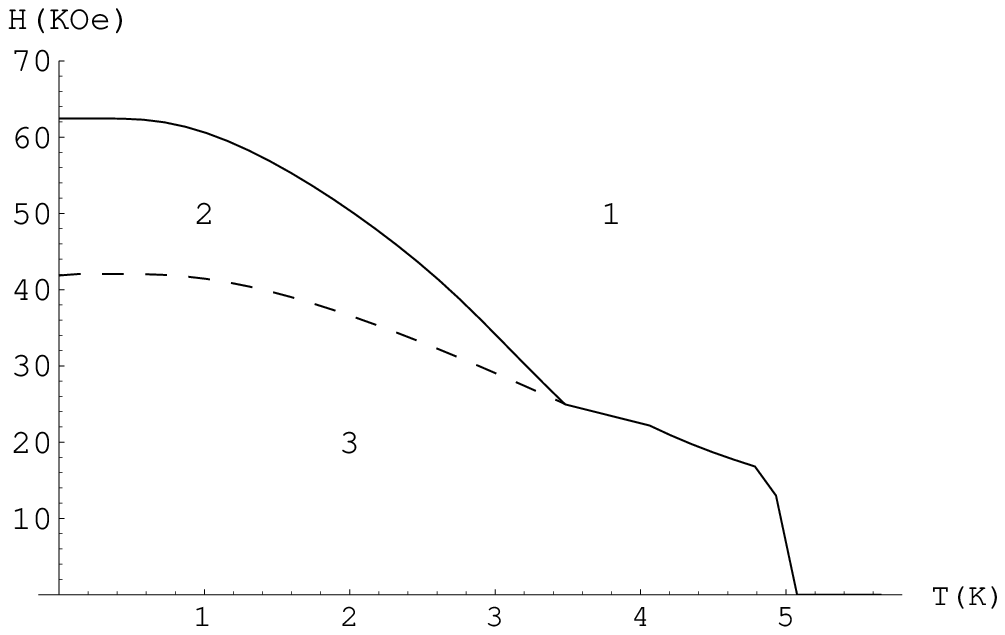,height=4.2 cm}
\caption{\label{fig2}Phase diagram for a wide-band AFS 
with carrier concentration $n=0.50\, n_{\text{T}}$} 
%\end{figure}

%\begin{figure}
\epsfig{file=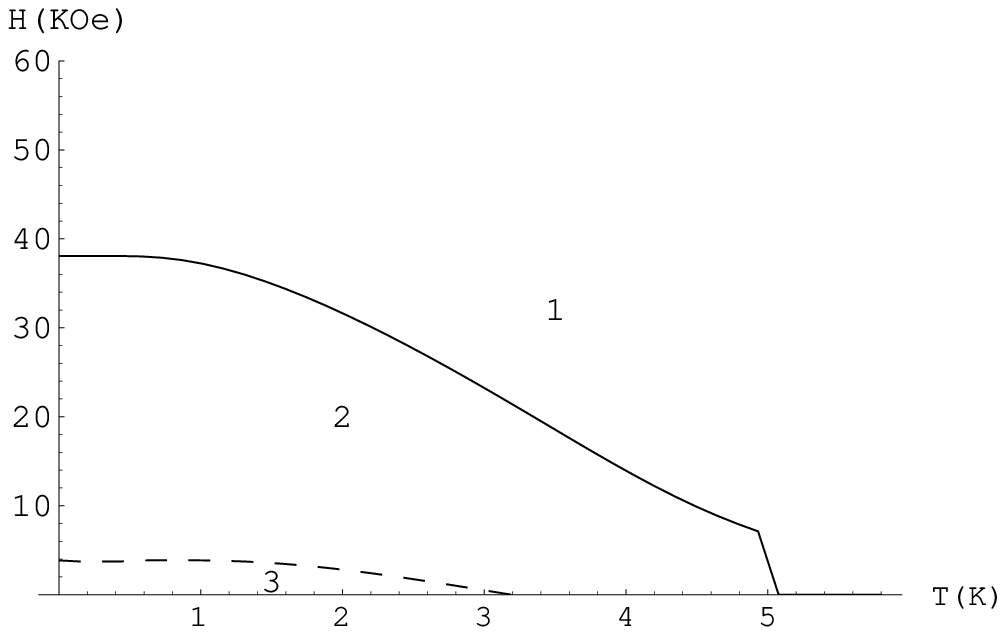,height=4.2 cm}
\caption{\label{fig3}Phase diagram for a wide-band AFS 
with carrier concentration $n=0.95\, n_{\text{T}}$} 
%\end{figure*}
\end{figure}

Phase diagrams for wide-band semiconductors are shown in figures       %
(\ref{fig1}), (\ref{fig2}) and (\ref{fig3}) for conduction electron
concentrations equal to $n=0.10,\,0.50,\,0.95\, n_{\text{T}}$,
respectively. The areas 1, 2, 3 denote regions of absolute stability
for the homogeneous state, the conductive phase-separated state and
the insulating phase-separated state, respectively. The parameters
that characterize these materials are chosen to be the same as in
reference \cite{Nag90} (they correspond to rare-earth compounds 
like europium chalcogenides): $S=\frac{7}{2}$, 
$|J|S=10^{-3}$ eV (implying $T_{\text{N}}=5$ K, and the field at
which both sublattices collapse is $98.7$ kOe), $AS=1$ eV, 
$\epsilon_{r}=20$, $a^{-3}=4\cdot10^{22}$ cm$^{-3}$, and the electron
effective mass is equal to free electron mass, this implies  $W=4$ eV. 
The value obtained for $n_{\text{T}}=1.10\cdot10^{20}$ cm$^{-3}$. 
The value for $n_{\text{u}}$, at which the conducting 
phase-separated state becomes unstable at $T=H=0$, is 
$1.79\cdot10^{20}$ cm$^{-3}$.

We see that at low carrier concentrations only a transition from the   %
insulating phase-separated state to the homogeneous state is possible
(see figure  (\ref{fig1})). The physical explanation of the diagram 
(\ref{fig1}) is the following. At fixed $n$, the effect of the increase
of magnetic field and temperature is to decrease the value of $x$ with
respect to its value at $T=H=0$. In the range of concentrations we are
working, $n_{\text{d}}<n<n_{\text{T}}$, the state of the sample at 
$T=H=0$ is an insulating phase-separated state ($x>1$). Regarding to 
the behavior of $x$ at $T=H=0$ with $n$, $x\rightarrow 1$ as 
$n\rightarrow n_{\text{T}}$. Following equations (\ref{ meanfield}),
the d-spins of the magnetic ions in the FM part of the phase-separated 
state experience a magnetic field which is the sum of the external 
field, $H$, and an additional field created by conduction electrons,
$H_{\text{e}}$. If we set the external field equal to zero, 
$H_{\text{e}}$ is large enough to establish the FM  order in the
multiply-connected part, i.e. $H_{\text{e}}>2JS_{1}$. But as we 
increase external field, the value of $x$ decreases very quickly and 
it is reached a point in which $H+\frac{An}{2}(1+x)=2JS_{1}$. On 
further increase of $H$, the magnetic field cannot maintain the 
multiply-connected part in a FM (NSFM) state and a transition to an
homogeneous AF (NCAF) state occurs. The transition line can be drawn
in a simple approximate way. We start by looking for the value of $x$,
$x_{0}$, which makes the energy minimal at $T=0$. For
$n=0.10\, n_{\text{T}}$ we found $x_0=2.525$. Because this value 
increases only slightly as we move along the transition line, this can
be approximated by the expression $H+\frac{An}{2}(1+x_0)=2JS_{1}$.

At $T=0$ for $n\geq0.18\, n_{\text{T}}$, the carrier concentration is  %
high enough to permit that increasing the external field we remain in
the insulating phase-separated state until $x=1$. Therefore for 
concentrations  $n>0.18\, n_{\text{T}}$, three different phases appear.
At low temperatures and low magnetic fields, we have a transition 
between both of the phase-separated states. This line correspond to the 
simultaneous percolation of the magnetic order and the conduction 
electron system. It is almost a second order transition. The values of
$x$ at both sides of the transition line are $x=1.09$ and $x=0.92$ at 
$T=0$ not varying with $n$, increasing as we move along the line
towards higher temperatures, and being the growth smaller, the closer
$n$ to $n_{\text{T}}$. To calculate the values of $T,H,x_{3},x_{2}$ 
(being $x_{3},x_{2}$ the values of $x$ at both sides of the transition
line), we impose three conditions: the free energy of both 
phase-separated states must have a minimum and the free energy of both
states must be equal. This allows us to calculate $H,x_{3},x_{2}$
varying $T$ along the line. Over the same range of temperatures but at
higher fields, we find a transition between conductive phase-separated 
and an homogeneous state. This is a first order transition, similar to
the melting of a solid. Again the same effect than in diagram 
(\ref{fig1}) appears once we are in the conductive phase-separated 
state, the value of $n$ must be high enough to maintain the 
simply-connected in a FM state. If not, a first-order transition to an
homogeneous state with a value of $x$ approximately constant along the
line occurs. 

In figure (\ref{fig2}), $n$ is high enough to permit that increasing   %
the magnetic field we remain in the conducting phase-separated state
until $x=0$, so this effect is absent. In this case the transition line
is determined in the same way as in the usual melting transition 
\cite{TomoV}. To calculate the value of $T,H,x_{2}$, we impose the 
conditions that the first two derivatives of the free energy of the 
conductive phase-separated state be equal to zero. At higher 
temperatures we find the transition line between the insulating 
phase-separated and homogeneous states, that now corresponds to a 
NSFM order. It is important to notice that what we denote as phase 1 
in the diagrams, corresponds only to homogeneous magnetic states, 
without distinguishing whether they are of the NCAF or NSFM type. 
This is the reason why a kink in the transition line appear at
$T\simeq5$ K. Really, it corresponds to a tri-critical point between
insulating phase-separated, homogeneous NSFM states (at lower 
temperatures) and homogeneous NCAF (at higher temperatures). The 
transition line between these two is not drawn.
The transition line from the kink to the $T$-axis correspond to the    
magnetic transition of the AF part of the insulating phase-separated 
state to a PM state. With $n\longrightarrow n_{\text{T}}$, this
tri-critical point tends to the point $T=T_{N},H=0$. Due to the
absence of electrons in the AF part of the phase-separated states the 
whole the sample has the same N\'eel temperature, as is confirmed by 
experiment. This part of the line is calculated in the same way that 
the transition line between conductive phase-separated state and
homogeneous states.

The third diagram corresponds to $n=0.95\, n_{\text{T}}$. We find two  %
transition lines. One between both of the phase-separated states, 
close to second order, as in the preceding discussion. We see that now
the tricritical point between conductive phase-separated, insulating
phase-separated and homogeneous states is absent. Therefore the
percolation with temperature (this is, at $H=0$) is also possible. The
other line corresponds to the transition between conductive
phase-separated and homogeneous states (now, NSFM all along the line).
The kink was moved to higher temperatures. We see that the region of
absolute stability of the insulating phase-separated state has been 
reduced, tending to dissappear as $n$ goes to $n_{\text{T}}$.   

\subsection{$W<<AS$}

\begin{figure}
%\begin{figure*}
\epsfig{file=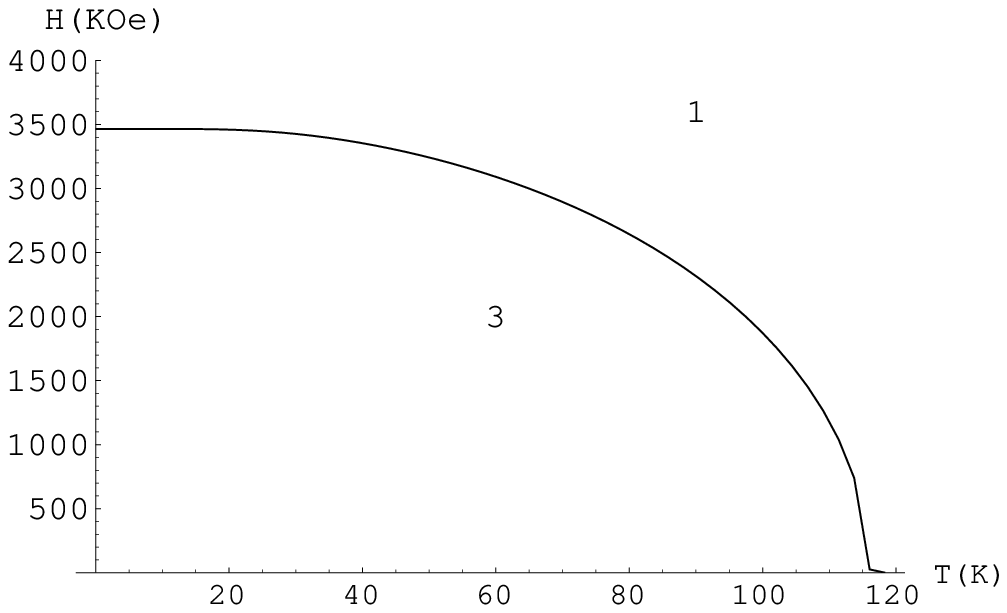,height=4.2 cm}
\caption{\label{fig4}Phase diagram for a ``double exchange''
AFS with carrier concentration $n=0.10\, n_{\text{T}}$}
%\end{figure}

%\begin{figure}
\epsfig{file=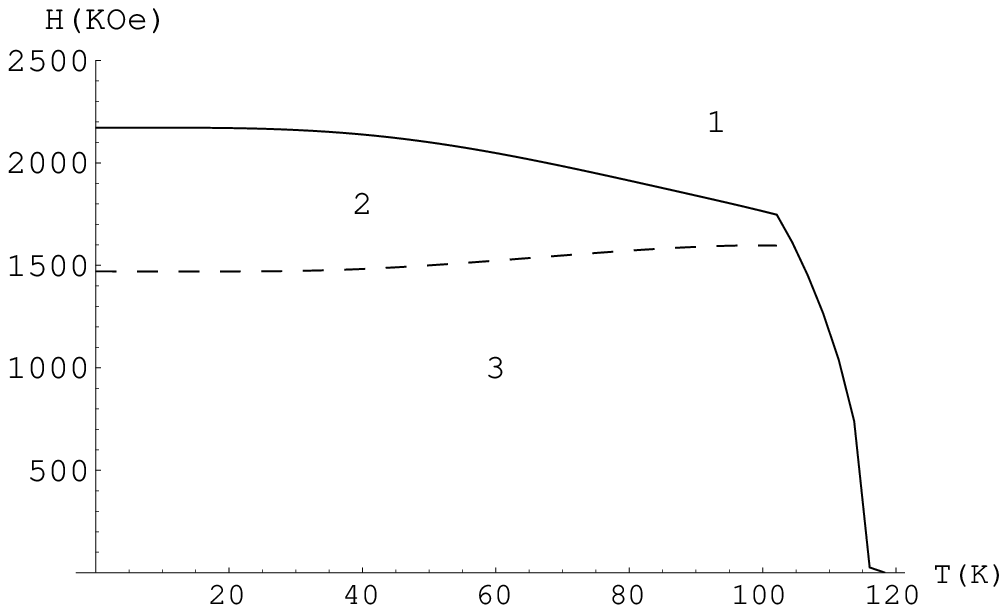,height=4.2 cm}
\caption{\label{fig5}Phase diagram for a ``double exchange''
AFS with carrier concentration $n=0.50\, n_{\text{T}}$} 
%\end{figure}

%\begin{figure}
\epsfig{file=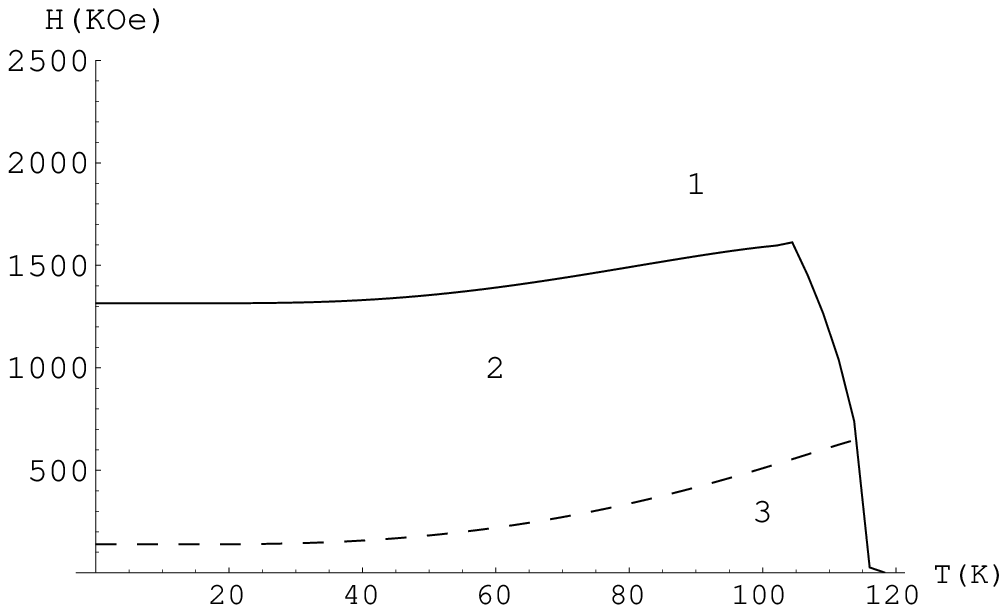,height=4.2 cm}
\caption{\label{fig6}Phase diagram for a ``double exchange''
AFS with carrier concentration $n=0.95\, n_{\text{T}}$} 
%\end{figure*}
\end{figure}

Phase diagrams for ``double-exchange'' semiconductors are              %
essentially the same. We also present three of them, in figures
(\ref{fig4}), (\ref{fig5}), and (\ref{fig6}), 
corresponding to conduction electron concentrations 
equal to $n=0.10,\,0.50,\,0.95\, n_{\text{T}}$, respectively. 
The parameters that characterize ``double-exchange'' semiconductors
are the following (typical for lanthanium manganites, for example)
\cite{Nag01}: $S=2$, $|J|S=2\cdot10^{-2}$ eV (that 
implies $T_{N}=116$ K, and the field at which the two sublattices 
collapse at $T=0$ is $3.46$ MOe), $\epsilon_{r}=5$, 
$a^{-3}=4\cdot10^{22}$ cm$^{-3}$, the effective electron mass equal 
to the free electron mass, that implies  $W=4$ eV and $A=4$ eV. With 
these parameters we obtain $n_{\text{T}}=9.16\cdot 10^{20}$ cm$^{-3}$
and $n_{\text{u}}=1.47\cdot10^{21}$ cm$^{-3}$.

The behavior of the ``double exchange'' compounds is quantitatively    %
different. The scenario discussed above for wide-band case remains
valid, but a higher value of $A$ implies a very strong magnetic field
$H_{\text{e}}$ acting on the FM part of the separate states and also
a higher value of $n_{\text{T}}$. It must be stressed that the values 
that we obtain for $n_{\text{T}}$, are far from the true theoretical
values. In particular, the values obtained depend on the physical 
parameters such as $A$, while we are dealing with a geometric 
transition that cannot depend on these details. This is because we 
lost the fractal nature of the surface that separes phases to simplify 
calculations. The quantum nature of the d-spin system forces the true
percolation concentration to be $n_{\text{per}}=0.16 a^{-3}$, the 
percolation concentration for the Bethe lattice (see, for example,
\cite{Efr94}), \textit{independently} of the physical processes 
involved. So, in both cases treated here, it is necessary to ``map''
$n_{\text{T}}\longrightarrow n_{\text{per}}$ in order to compare with 
the experimental values.  
Because $H_{\text{e}}S>>T_N$ we can neglect the terms with $m\neq S$ 
in the partition function of the FM part. This makes the FM part 
temperature-independent, remaining in the same state as at $T=0$.  
We see in figures (\ref{fig5}) and (\ref{fig6}), that the tri-critical
point corresponding to the kink at $T\simeq100$ K, is now between 
conductive phase-separated, homogeneous NSFM (at lower temperatures)
and homogeneous NCAF (at higher temperatures), and appears before in 
temperature than the tri-critical point between conductive 
phase-separated, insulating phase-separated and homogeneous state 
(now, NCAF). As we see in figure (\ref{fig6}),
percolation is possible varying the magnetic field in the same way as
above, but percolation in temperature (i.e. at $H=0$) is not, contrary 
to what was seen above. However, this result can be modified beyond a 
mean field approximation. It must also be stressed that the value of 
$AS$ sets the scale of the $H$-axis and its high value makes the
transition to the homogeneous states occur to unphysical values of 
the external field.

A comparison with the experimental results is specially relevant for   %
the case of ``double-exchange'' compounds. It must be stressed that
although all along this article we talk about the charge carriers as
being conduction electrons, the model is general enough to deal also
with hole-doped compounds. An example of the latter could be the
canonical CMR lanthanium manganite La$_{1-x}$Ca$_{x}$MnO$_{3}$ 
with $x<0.16$, which is extensively reviewed in the reference
\cite{Nag01R}. However the present calculation is more interesting for
electron-doped  compounds, such as Sm$_{1-x}$Ca$_{x}$MnO$_{3}$ with
$x>0.84$, because the region of record negative magnetoresistance fall
within the range of doping studied in this work. This compound is
studied in the reference \cite{Alg02}, where the observed CMR effects 
are explained as consequence of the metal-insulator transition 
described here. They also present the phase diagram in the $T-H$ plane 
for th compound Sm$_{0.15}$Ca$_{0.85}$MnO$_{3}$, which agrees
qualitatively and quantitatively with our figure (\ref{fig6}).

\begin{acknowledgments}
The authors are deeply indebted to Prof. E. L. Nagaev for his
encouragement and help during its realization.
Unfortunately, Prof. Nagaev died recently. This article is dedicated 
to his memory.
\end{acknowledgments}

\bibliography{article2}
\end{document}